\documentstyle[12pt,epsfig]{article}
 \newcommand{\be}{\begin{eqnarray}}
 \newcommand{\ee}{\end{eqnarray}}
 \newcommand{\beq}{\begin{equation}}
 \newcommand{\eeq}{\end{equation}}
 \newcommand{\ba}{\begin{array}{1}}
 \newcommand{\ea}{\end{array}}
 \newcommand{\bb}{\begin{thebibliography}}
 \newcommand{\eb}{\end{thebibliography}}

 \newcommand{\abstitle}[1]{{\small {\bf #1}}}
 \newcommand{\absauthor}[1]{{\small {\bf #1}}}
 \newcommand{\address}[1]{{\it #1}}
 
\begin{document}
 \begin{center}
 \abstitle{Gluon distribution in proton at soft and hard pp collisions}\\
 \vspace{0.3cm}
 \absauthor{G.I.Lykasov$^{1,4}$, V.A.Bednyakov$^1$, A.A.Grinyuk$^1$,   
M.Poghosyan$^{2,3}$\\
 and A.G.Dolbilov$^1$} \\ [2.0mm]
 \address{$^1$ JINR, Dubna, Moscow region, 141980,  Russia, \\ 
$^2$ Torino University, Torino, Italy\\
$^3$ CERN, Geneva, Switzerland \\
$^4$ E-mail:~lykasov@jinr.ru
}
 \end{center}
 \vspace{0.1cm}
 \vspace{0.2cm} 
{\bf Abstract}
 \vspace{0.1cm}

We analyze the inclusive spectra of hadrons produced in $pp$ collisions at
high energies in the mid-rapidity region within the soft QCD and perturbative 
QCD assuming the possible creation of the soft gluons at low intrinsic transverse 
momenta $k_t$. From the best description of the LHC data we found the parametrization
of the unintegrated  gluon distribution  which at low $k_t$ is different from the one
obtained within the perturbative QCD.

\section{Introduction}
\label{1}
As is well known, hard processes involving incoming protons, such as deep-inelastic 
lepton-proton
scattering (DIS), are described using the scale-dependent PDFs.
A distribution like this is usually calculated as a function of the Bjorken variable $x$
and the square of the four-momentum transfer $q^2=-Q^2$, integrated over the parton transverse 
momentum $k_t$.
However, for semi-inclusive processes, such as inclusive jet production in DIS,
electroweak boson production \cite{Ryskin:2003}, etc., the parton distributions unintegrated 
over  
the transverse momentum $k_t$ 
are more appropriate. The theoretical analysis of the unintegrated quark and gluon PDFs can 
be found, for example, in \cite{GBW:99}-\cite{Jung:04}.
According to  \cite{Ryskin:2010}, the gluon 
distribution function 
$g(k_t)$ at fixed $Q^2$ has the very interesting behaviour at small $x\leq 0.01$, it increases
very fast starting from almost zero values of $k_t$.  
In contrast, the quark 
distribution $q(k_t)$ is almost 
constant 
in the whole region of $k_t$ up to $k_t\sim$ 100 GeV$/c$ and much smaller
than $g(x)$.
This parametrization of the PDFs was obtained in \cite{Ryskin:2010} within the leading order 
(LO) and next-to-leading order 
of QCD (NLO) 
at large $Q^2$ 
from the known
(DGLAP-evolved \cite{{DGLAP}}) parton densities determined from the global data analysis. 
At small values of $Q^2$ the nonperturbative effects should be included to get the PDFs.
The nonperturbative effects can arise from the complex structure of the QCD vacuum.  
For example, 
within the instanton approach \cite{Kochelev:1998} the very fast increase of the unintegrated 
gluon distribution function at $0\le k_t\le 0.5$ GeV$/c$ and $Q^2=1$ (GeV$/c$)$^2$ is also shown.
These results stimulated us to assume, that the unintegrated gluon distribution in the 
proton can be included by
analyzing also the soft hadron production in $pp$ collisions.  
In this paper we analyze the inclusive spectra of the hadrons produced in $pp$
collisions at LHC energies in the mid-rapidity region including the possible creation 
of soft gluons in the proton. 
We estimate the unintegrated gluon distribution function (UGDF) at low intrinsic transverse
momenta $k_t\leq 1.5-1.6$ GeV$/$c and its parameters extract from the best description of the
LHC data at low transverse momenta $p_t$ of the produced hadrons. We also show that our UGDF
is similar to the UGDF obtained in \cite{GBW:99,Jung:04} at large values of $k_t$. 
\section{Inclusive spectra of hadrons in $pp$ collisions}
\label{sec:1}
\subsection{Quark-gluon string model (QGSM)}
Let us analyze the hadron production in $pp$ collisions
within the QGSM \cite{kaid1} or the dual parton model (DPM) \cite{capell2}
 including the transverse 
motion of quarks and diquarks in colliding protons \cite{LS:1992,BLL:2010}. As is
 known, the cylinder-type graphs
presented in Fig.1 make the main contribution to this process. 
 The left diagram of Fig.1, the so-called one-cylinder graph, 
corresponds to the cut one-pomeron graph and the 
right diagram of Fig.1, the so-called multi-cylinder graph, corresponds 
 to the cut n-pomeron graphs.
\begin{figure}[h!]
\centerline{\includegraphics[width=0.8\textwidth]{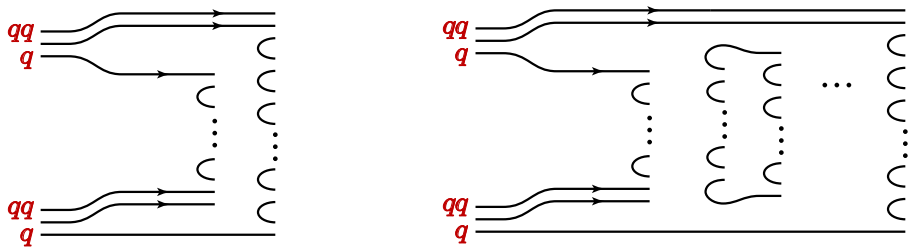}}
\caption{The one-cylinder graph (left) and the multi-cylinder graph (right)}
\label{fig:1}       
\end{figure}
The general form for the invariant inclusive hadron spectrum within the QGSM is the following 
\cite{kaid1}:
\be
\rho(x,p_t)\equiv E\frac{d\sigma}{d^3{\bf p}}
=
\sum_{n=1}^\infty \sigma_n(s)\phi_n(x,p_t)~,
\label{def:invsp}
\ee
where $E,{\bf p}$ are the energy and three-momentum of the produced hadron $h$ in the l.s.
 of colliding
protons respectively; 
$x,p_t$ are the Feynman variable and the transverse momentum of $h$; $\sigma_n$ is the
 cross
section for production of the $n$-pomeron chain (or $2n$ quark-antiquark strings) decaying
 into hadrons, 
calculated within the ``eikonal approximation'' \cite{Ter-Mart}
the function
$\phi_n(x,p_t)$ has the following form \cite{LS:1992}: 
\be
\phi_n(x,p_t)=\int_{x^+}^1dx_1\int_{x_-}^1 dx_2\psi_n(x,p_t;x_1,x_2)~,
\label{def:phin}
\ee
where
\be
\psi_n(x,p_t;x_1,x_2)=
\nonumber \\
F_{qq}^{(n)}(x_+,p_t;x_1)F_{q_v}^{(n)}(x_-,p_t;x_2)/F_{q_v}^{(n)}(0,p_t)+~
\nonumber \\
+ F_{q_v}^{(n)}(x_+,p_t;x_1)F_{qq}^{(n)}(x_-,p_t;x_2)/F_{qq}^{(n)}(0,p_t)+~
\nonumber \\
2(n-1)F_{q_s}^{(n)}(x_+,p_t;x_1)F_{{\bar q}_s}^{(n)}(x_-,p_t;x_2)/F_{q_s}^{(n)}(0,p_t)~.
\label{def:psin}
\ee
and $x_{\pm}=0.5(\sqrt{x^2+x_t^2}\pm x), x_t=2\sqrt{(m_h^2+p_t^2)/s}$,
\be
F_\tau^{(n)}(x_\pm,p_t;x_{1,2})=
\nonumber \\
\int d^2k_t{\tilde f}_\tau^{(n)}(x_\pm,k_t){\tilde G}_{\tau\rightarrow h}
\left(\frac{x_\pm}{x_{1,2}},{\tilde k}_t;p_t)\right)~,
\label{def:Ftaux}
\ee
\be
F_\tau^{(n)}(0,p_t)=
\nonumber \\
\int_0^1 dx^\prime d^2k_t {\tilde f}_\tau^{(n)}(x^\prime,k_t)
{\tilde G}_{\tau\rightarrow h}(0,p_t)={\tilde G}_{\tau\rightarrow h}(0,p_t)~.
\label{def:Ftauzero}
\ee
Here $\tau$ means the flavour of the valence (or sea) quark or diquark, 
${\tilde f}_\tau^{(n)}(x^\prime,k_t)$
is the quark distribution function depending on the longitudinal momentum fraction $x^\prime$
 and the 
transverse momentum $k_t$ in the $n$-pomeron chain; 
${\tilde G}_{\tau\rightarrow h}=z{\tilde D}_{\tau\rightarrow h}(z,{\tilde k}_t;p_t)$, 
${\tilde D}_{\tau\rightarrow h}(z,{\tilde k}_t;p_t)$ is 
the fragmentation function of a quark (antiquark) or diquark of flavour $\tau$ into
 a hadron $h$. 
\subsection{Hadron production in the mid-rapidity region}.
According to the Abramovskiy-Gribov-Kancheli cutting rules (AGK) \cite{AGK}, 
at mid-rapidity only Mueller-Kancheli type diagrams contribute to the inclusive 
spectrum of hadrons. 
It was shown in \cite{LS:1992,BGLP:2011} that at $x=0$ the function 
$F_\tau^{(n)}(x_\pm\simeq 0,p_t,x_{1,2})$ does not depend on $n$. Therefore, the function 
$\phi_n(0,p_t)$ becomes proportional to $n$.
In accordance with \cite{BGLP:2011}, the inclusive spectrum corresponding to the
contribution of quarks at $x=0$ has the following simple form:
\be
\rho_q(0,p_t)=
{\tilde\phi}_q(0,p_t)\sum_{n=1}^\infty n \sigma_n(s),
\nonumber \\
g(s/s_0)^{\Delta}{\tilde\phi}_q(0,p_t),
\label{def:invspq}
\ee 
where ${\tilde\phi}_q(0,p_t)$ is the so-called vertex function in the Mueller-Kancheli
 type diagram which is determined 
by the density distribution of the particles produced in the central region;
$\Delta=0.12$, g$=$21 mb and $\sigma_{nd}$ is the nondiffractive cross section due
to the exchange of any number cut-pomerons.

According to Eqs.(\ref{def:phin}-\ref{def:psin}), the sea quarks contribute to $\phi_n$
 and the inclusive spectrum
at $n\geq 2$. Assuming possible creation of soft gluons in the proton, which are split
 into $q{\bar q}$ pairs and 
should vanish at the zero intrinsic transverse momentum ($k_t\sim 0$), one can obtain
 the following energy dependence 
for the gluon contribution $\rho_g(0,p_t)$ to the inclusive spectrum \cite{BGLP:2011}:
\be
\rho_g(x=0,p_t)=\sum_{n=2}^\infty \sigma_n(s)\phi_n^g(0,p_t)\equiv 
\nonumber \\
(g(s/s_0)^{\Delta}-\sigma_{nd})
{\tilde\phi}_g(0,p_t)~,
\label{def:invspg}
\ee
Finally, we can present the inclusive spectrum at $x\simeq 0$ in the following form:
\be
\rho(p_t)=\rho_q(x=0,p_t)+\rho_g(x=0,p_t)~. 
\label{def:rhoagk}
\ee    
The functions ${\tilde\phi}_q(0,p_t)$ and ${\tilde\phi}_g(0,p_t)$ are fixed from the fit of the
experimental data on the inclusive spectra of the charged hadrons produced in the central
 $pp$ collisions
at different initial energies. 
So, we found the following parametrization for ${\tilde\phi}_q(0,p_t)$ and
 ${\tilde\phi}_g(0,p_t)$ \cite{BGLP:2011}:   
\be
{\tilde\phi}_q(0,p_t)=A_q\exp(-b_q p_t)~
\nonumber \\
{\tilde\phi}_g(0,p_t)=A_g\sqrt{p_t}\exp(-b_g p_t),
\label{def:phiq}
\ee
 charged particles at $y=0$:
where  
$A_q=4.78\pm 0.16$ (GeV$/$c)$^{-2}$,~$b_q=7.24\pm 0.11$ (Gev/c)$^{-1}$ and
 $A_g=1.42\pm 0.05$ (GeV$/$c)$^{-2}$;~ 
$b_g=3.46\pm 0.02$ (GeV/c)$^{-1}$.
{
\subsection{Gluon distribution in proton within GBW model}
The conventional QGSM or DPM model does not include the distribution of gluons in the proton.
 However,   
as is well known, at large transverse momenta $p_t$ of hadrons the gluons in the proton play
very important role in description of the
experimental data. Therefore, one can assume that the contribution of the gluon 
distribution in the proton to the inclusive spectrum 
of the produced hadrons slowly appears when $p_t$ increases and it will be  sizable 
at high values
 of $p_t $. This assumption is 
also confirmed by the increase of the UGDF in the proton at
 $x\sim 0$ as a function of the 
internal transverse 
momentum $k_t$ when $k_t$ grows \cite{GBW:99}-\cite{Jung:04}.

According to Refs.\cite{GBW:99,Jung:04}, the UGDF,
as a function of $k_t$ at some value of $Q_0$ and low $x$ is presented in the following 
form:
\be
xg(x,k_t,Q_0)=
C_0R_0^2(x)k_t^2
\exp\left(-R_0^2(x)k^2_t\right)~,
\label{def:GBWgl}
\ee 
where $C_0=3\sigma_0/\left(4\pi^2\alpha_s(Q_0)\right)$,~ $R_0(x)=$GeV$^{-1}(x/x_0)^{\lambda/2}$, 
$x_0,\lambda$ and $\sigma_0$ are defined in \cite{GBW:99,Jung:04}; $\alpha_s(Q_0)$ is
the QCD coupling constant. 
To get the UGDF at low $k_t$ we assume the possible creation of the soft gluons in the proton
which appears at nonzero $k_t$. 
We calculated the gluon contribution  ${\tilde\phi}_g(0,p_t)$ entering into 
Eq.(\ref{def:invspg}) as 
the cut graph of the one-pomeron exchange in the gluon-gluon interaction (Fig.2, right) 
using the splitting of the gluons into the $q{\bar q}$ pair. Then the calculation was 
made in a way similar to the calculation of the sea quark contribution to the inclusive 
spectrum within the QGSM, see Eq.(\ref{def:psin}) at $n=2$. 
\begin{figure}[h!]
\centerline{\includegraphics[width=0.8\textwidth]{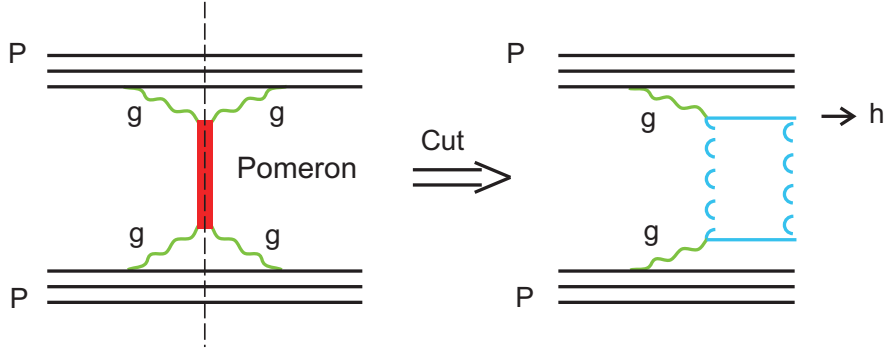}}
\caption 
    {The one-pomeron exchange graph between two gluons in the elastic $pp$ scattering (left) 
and the corresponding cut graph (right).} 
\label{Fig_1}
\end{figure} 
Calculating the diagram of Fig.2 (right) we assumed the following form for the $xg(x,k_t,Q_0)$:
\be
xg(x,k_t,Q_0)= 
C_0 C_1 (1-x)^{b_g}\times
\nonumber \\
\left(R_0^2(x)k_t^2+C_2(R_0(x)k_t)^a\right)
\exp\left(-R_0(x)k_t\right)~,
\label{def:gldistr}
\ee 
where 
 $C_2,a,b_g$ and 
$\lambda$ are the parameters which were found from the best description of data.
As it will be shown in the next section, the parametrization given by Eq.(\ref{def:gldistr})
results in the satisfactory description of the inclusive spectra of hadrons produced in $pp$
collisions at the mid-rapidity region and low transverse momenta $p_t\leq 1.5$ GeV$/$c. At higher values
of $p_t$ the gluon contribution ${\tilde\phi}_g(0,p_t)$ given by Eq.(\ref{def:phiq}) is decreasing very
fast and, in fact, does not contribute to the hadron spectrum at large $p_t$. Therefore, one can find 
the parametrization for the UGDF that is very close to the GBW parametrization  \cite{GBW:99,Jung:04}
at large $k_t$ and coincides with the UGDF given by Eq.(\ref{def:gldistr}) at low $k_t$. It is the following: 
\be
xg(x,k_t,Q_0)= 
C_0 C_3 (1-x)^{b_g}\times
\nonumber \\
\left(R_0^2(x)k_t^2+C_2(R_0(x)k_t)^a\right)
\exp\left(-R_0(x)k_t-d(R_0(x)k_t)^3\right)~,
\label{def:gldistrnew}
\ee 
The coefficients $C_1,C3$ are found from the normalization
\be
xg(x,Q_0^2)=\int_0^{Q_0^2} dk_t^2 xg(x,k_t,Q^2)
\label{def:normint}
\ee
\section{Results and discussion}
The results ~ of our calculations of ~ the charged ~ hadron ~ inclusive
spectrum $(1/N_{ev})d^3N/dydp_t^2$, 
where $N_{ev}$ is the total number of the inelastic events, 
in the mid-rapidity at $\sqrt{s}=$7 TeV are presented in Fig.(3).
The solid curve corresponds to the quark contribution $\rho_q(x=0, p_t)$ given by
Eq.(\ref{def:invspq}), the long-dash line is the gluon contribution $\rho_g(x=0, p_t)$
 (\ref{def:invspg}) to the inclusive spectrum; 
the dotted curve corresponds to the sum of both contributions, see 
Eqs.(\ref{def:rhoagk}).
One can see that the conventional quark contribution $\rho_q(y=0,p_t)$ 
is able to describe the data up to $p_t\leq$1 GeV$/c$, whereas the inclusion of the gluon
 contribution allows 
us to extend the range of good description up to 2 GeV$/c$. At larger values of $p_t$ the 
contribution of hard processes is not negligible.
Therefore, we calculated the inclusive spectrum at $p_t\geq$ 2 GeV$/$c 
within the leading order perturbative QCD (LO PQCD) \cite{AVEF:1974}-\cite{FF:AKK08}, 
see the details
in \cite{BGLP:2011}. The short-dash curve in Fig.3 (top) corresponds to our LO PQCD.
    \begin{figure}[h!]
\centerline{\includegraphics[width=0.6\textwidth]{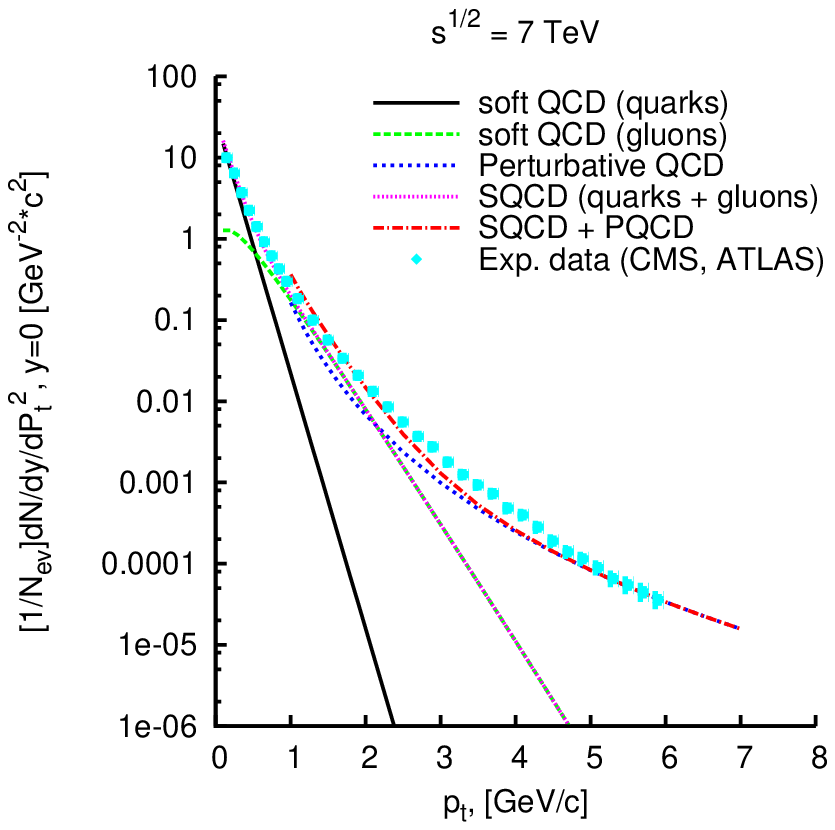}}
\centerline{\includegraphics[width=0.6\textwidth]{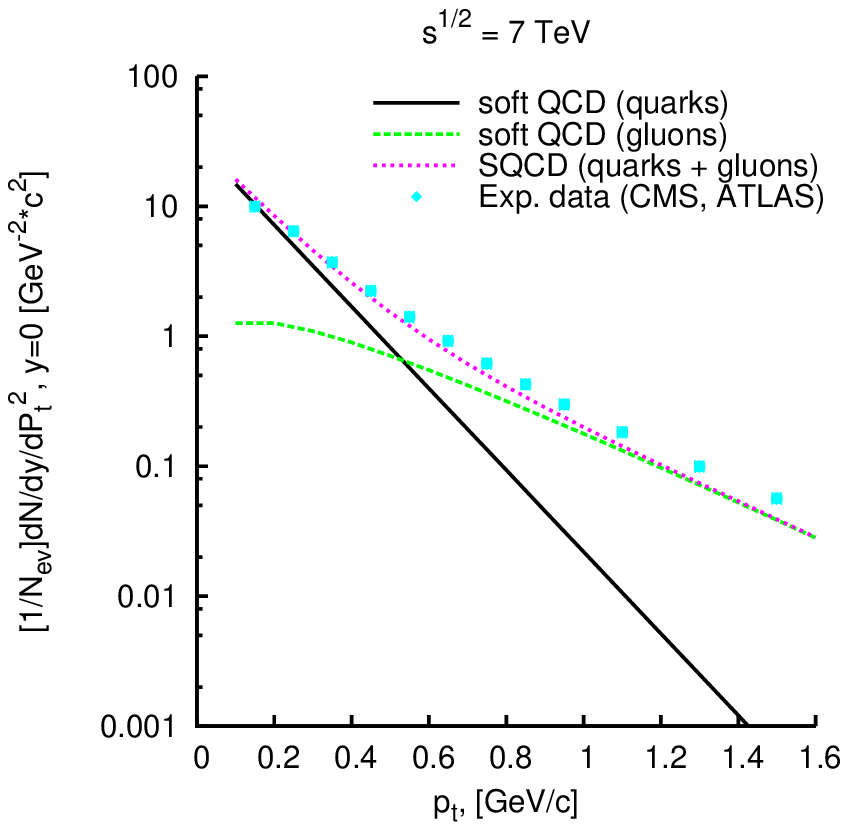}}
  \caption[Fig.2]{The inclusive spectrum of charged hadron as a function of $p_t$ (GeV$/c$)
in the central rapidity region ($y=0$) at $\sqrt{s}=$7 TeV and wide region of
$p_t$ (top) and the same spectrum at $p_t\leq$ 1.6 GeV$/$c (bottom)
 compared with the CMS \cite{CMS} which are very close to the
ATLAS data..}
  \end{figure} 
\begin{figure}[h!]
\centerline{\includegraphics[width=1.0\textwidth]{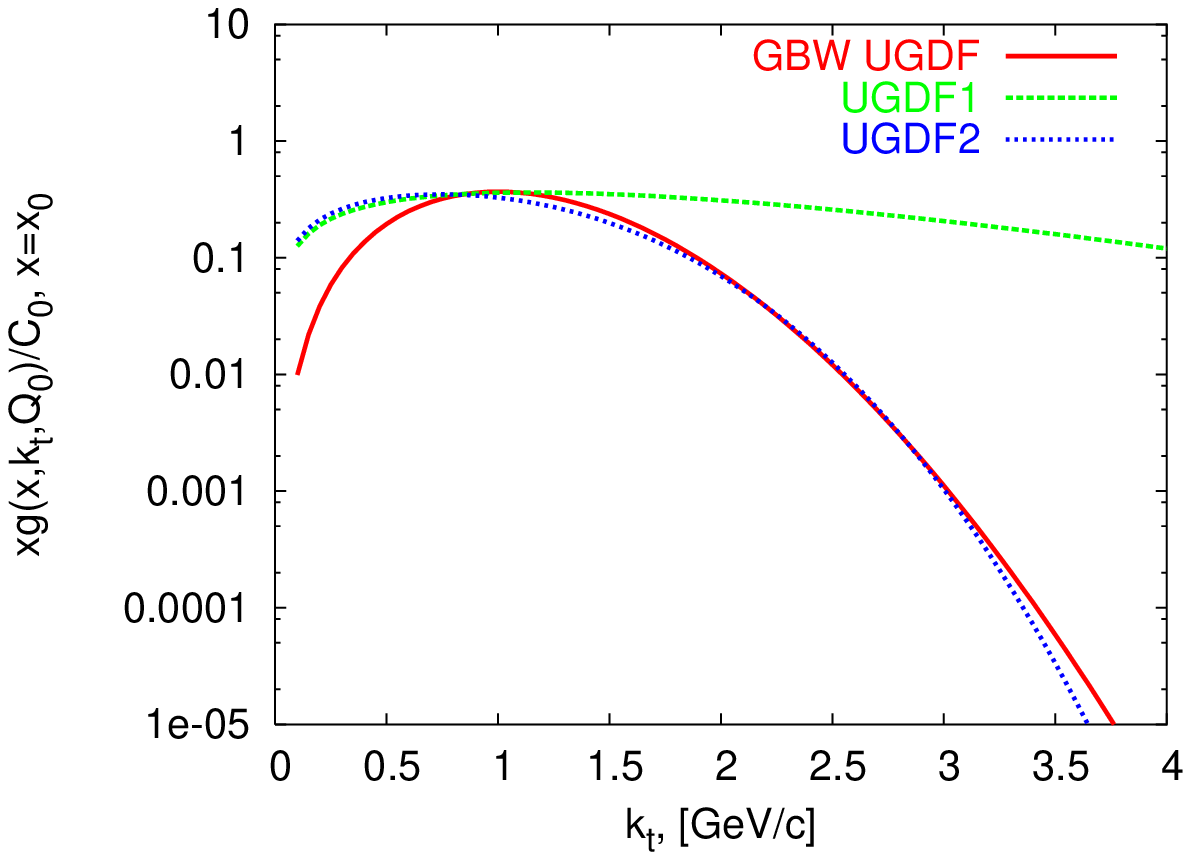}}
\caption 
    {The unintegrated  gluon distribution $xg(x,k_t,Q_0)/C_0$ as a function of $k_t$
at $x=x_0$ and $Q_0=1.$GeV$/$c. The solid line is the GBW UGDF  
\cite{GBW:99,Jung:04}, the dashed curve (UGDF1) corresponds to Eq.(\ref{def:gldistr})
and the dotted line (UGDF2) corresponds to Eq.(\ref{def:gldistrnew}) which is very close to 
the GBW UGDF at $k_t\geq 1.5$ GeV$/$c.
} 
\label{Fig_1}
\end{figure} 
From the best fit of the inclusive $p_t$-spectra of the charged hadrons produced in  
$pp$ collisions at $x\simeq 0$ presented in Fig.3 (bottom) by the short dash line
we found the following parameter values entering 
into the form for $xg(x,k_t,Q_0)$ given by Eq.(\ref{def:gldistr}):
$$
a=0.7; C_2\simeq 2.3; \lambda=0.22; b_g=12; d=0.2; C_3=0.3295
$$\\
and
$$
C_1=\frac{2\pi}{\Gamma(4)}\left(1+C_2\Gamma(a+2)/\Gamma(4)\right)^{-1}~.
$$
In Fig.4 the UGDF $xg(x,k_t,Q_0)/C_0$ as a function of $k_t$
at $x=x_0$
and $Q_0=1.$GeV$/$c is presented. 
The solid line (GBW UGDF) is the GBW parametrization 
\cite{GBW:99,Jung:04} given by Eq.(\ref{def:GBWgl}) divided by $C_0$; 
the dashed curve (UGDF1) corresponds to the UGDF given by  Eq.(\ref{def:gldistr})
and the dotted line (UGDF2) corresponds to the UGDF given by  Eq.(\ref{def:gldistrnew}) 
which is very close to the GBW UGDF at $k_t\geq 1.5$ GeV$/$c. It is seen from Fig.4 
that the parametrization UGDF2 is different from the GBW UGDF mainly at 
$k_t\leq 1.4$ GeV$/$c.
\section{Conclusion} 
We assume that the contribution of the gluon
distribution in the proton to the inclusive spectrum 
of the produced hadrons slowly appears when $p_t$ increases and it will be  sizable at
high values of $p_t $. This assumption is 
also confirmed by the increase of the unintegrated gluon distribution in the proton
 at $x\sim 0$ as a function of the internal transverse 
momentum $k_t$ when $k_t$ grows \cite{Ryskin:2010,Kochelev:1998,GBW:99,Jung:04}.

Therefore, to illustrate this hypothesis we fit the experimental data on the inclusive
 spectra of charged particles
produced in the central $pp$ collisions at energies larger than the ISR starting
by the sum of
the quark contribution $\rho_q$ given by Eq.(\ref{def:invspq}) and the gluon
 contribution $\rho_g$ 
(see Eq.(\ref{def:invspg})).
 The parameters of this fit do not depend on the initial energy in that energy interval.
From the best fit of the LHC data on the inclusive spectra of the charged hadrons
produced in the mid-rapidity pp collisions at low $p_t$ we found a new parametrization
of the unintegrated gluon distribution in the proton at small values of the intrinsic
momentum $k_t$, which is similar to the UGDF obtained in \cite{GBW:99,Jung:04} at large
$k_t$.  
The inclusion of the hard $pp$ collision within LO QCD allows us to describe rather
 satisfactorily the data at higher
values of $p_t$.     
\vspace{0.2cm}
{\bf Acknowledgements}\\
The authors are very grateful to A.Bakulev, V.Cavazini, A.Dorokhov, A.V.Efremov, F.Francavilla,~ 
C. Gwenlan, H.Jung, V.Kim, ~B.Kniehl, N.Kochelev, E.A.Kuraev, A.V.Lipatov, L.N.Lipatov, ~T.Lomtadze, 
M.Mangano, ~C.Merino, S.V.Mikhailov, E.Nurse, F.Palla, E.Pilkington, C.Royon, M.G.Ryskin, ~ E.Sarkisyan-Grinbaum, 
O.V.Teryaev, Yu.Shabelski and N.P.Zotov for very useful discussions and comments. This work was supported 
in part by the Russian Foundation for Basic Research, project No:11-02-01538-a.


\end{document}